\documentclass[prl,twocolumn,
nofootinbib,preprintnumbers]{revtex4}

\usepackage{times}
\usepackage{natbib}
\usepackage{amssymb,amsbsy,amsmath,amsfonts}
\usepackage{mathrsfs}
\usepackage{graphicx}
\usepackage{epsf,epsfig,float,latexsym,amsthm,fancyhdr,rotating}
\usepackage{graphics,psfrag,longtable}
\usepackage{slashed}
\usepackage{esint}
\usepackage{nicefrac}

\usepackage[colorlinks,citecolor=blue,linktoc=all,linkcolor=cyan]{hyperref} 

\DeclareMathAlphabet{\mathantt}{OML}{antt}{l}{it}
\DeclareMathAlphabet{\mathpzc}{OT1}{pzc}{m}{n}

\def\beq{\begin{equation}}
\def\eeq{\end{equation}}
\def\bea{\begin{eqnarray}}
\def\eea{\end{eqnarray}}
\def\beqa{\begin{equation}\begin{array}{l}}
\def\eeqa{\end{array}\end{equation}}
\def\eqlab#1{\label{eq:#1}}
\def\figlab#1{\label{fig:#1}}

\def\Eqref#1{Eq.~(\ref{eq:#1})}

\def\Figref#1{Fig.~\ref{fig:#1}}

\def\boxfrac#1#2{\mbox{\small{$\frac{#1}{#2}$}}}
\def\half{\mbox{\small{$\frac{1}{2}$}}}

\def\barr{\left(\begin{array}{c}}
\def\earr{\end{array}\right)}
\def\bmat{\left(\begin{array}{cc}}
\def\emat{\end{array}\right)}
\def\al{\alpha}

\def\ga{\gamma} 
\def\de{\delta}

\def\nn{\nonumber}
\def\dd{\mathrm{d}}
\def\cO{\mathcal{O}}

\DeclareMathOperator\im{Im}
\DeclareMathOperator\re{Re}\def\3d{3-D}

\begin{document}
\title {Polarizability relations across real and virtual Compton scattering processes}

\author{Vladimir Pascalutsa}
\author{Marc Vanderhaeghen}

\affiliation{
Institut f\"ur Kernphysik, Cluster of Excellence PRISMA,  Johannes Gutenberg-Universit\"at, D-55099 Mainz, Germany}

\begin{abstract}
We derive two relations involving spin polarizabilities of a spin-1/2 particle and consider their empirical implications for the proton. Using the empirical 
values of the proton anomalous magnetic moment, electric and magnetic charge radii, 
moments of the spin structure functions
$g_1$, $g_2$, and of two spin polarizabilities, the present relations constrain the low-momentum behavior of 
generalized polarizabilities appearing in virtual Compton scattering. In the case of the proton, the dispersive
model evaluations of the spin and generalized polarizabilities appear to be consistent
with these relations. The ongoing measurements of different electromagnetic observables at the MAMI, Jefferson Lab, and HI$\gamma$S facilities may be able to put these relations
 to a test, or use them to unravel the low-energy spin structure of the nucleon.  
\end{abstract}
\date{\today}
\maketitle

Light absorption and scattering are long-known to be related ---
the Kramers--Kronig relation~\cite{KKrelation} and Baldin sum rule~\cite{Baldin}
providing the celebrated examples.
When the target particle (absorber/scatterer) 
has spin, the polarization of both the photon and the target may become
involved, leading to the celebrated Gerasimov--Drell--Hearn (GDH) sum rule \cite{Gerasimov:1965et,Drell:1966jv} 
and a lesser-known sum rule for the forward spin polarizability $\ga_0$
\cite{GellMann:1954db}. 
These
sum rules are derived from the general properties of analyticity,  unitarity, Lorentz and gauge symmetries of the forward Compton scattering (CS) amplitude. 
An extension to forward doubly-virtual Compton scattering (VVCS),
$\gamma^\ast p \to \gamma^\ast p$, allows one
to relate the momentum transfer ($Q^2$) dependencies of the absorption and elastic scattering processes \cite{Anselmino:1988hn, Ji:1999mr}, as well as to introduce longitudinal polarizabilities, such as $\de_{LT}$~\cite{Drechsel:2000ct,Drechsel:2002ar, Drechsel:2004ki}, 
which have been the subject of dedicated experimental activities at the Jefferson Lab, 
see~\cite{Kuhn:2008sy, Chen:2010qc} for reviews. 
These relations play an instrumental role in the studies of nucleon structure, where they provide model-independent constraints for the determination
of the various nucleon polarizabilities, see~\cite{Drechsel:2002ar}
for a review. 

Until now the generalized polarizabilities (GPs) \cite{Guichon:1995pu,Guichon:1998xv} appearing in the single virtual Compton scattering (VCS) process, 
$\gamma^\ast p \to \gamma p$, were escaping these constraints,
mainly because the apparent crossing asymmetry of this process
precludes any relation of the absorptive part 
of the VCS amplitude to an absorption cross section. 
The purpose of this letter is to show 
how an extension of the GDH sum rule to finite $Q$
may involve these GPs measured in the low-energy VCS process. 

We present first such sum rules, valid for a spin-$\nicefrac{1}{2}$ target, cf.\ Eq.~(\ref{qsqrgdhsr})
below. They involve two of the GPs: $P^{ (L1, L1)1}$ and $P^{(M1, M1)1}$,
where the superscript indicates the multipolarities - $L1(M1)$ denoting electric (magnetic) dipole 
transitions respectively - 
of the initial and final photons. Furthermore, the superscript $``1"$ denotes that these GPs represent two 
of the four lowest order spin-dependent GPs, involving a spin flip of the nucleon target corresponding to a change of angular momentum by one unit.
These GPs, accessed in the VCS process, are functions of the initial photon c.m.\ 3-momentum squared $\vec q^{\, 2}$, see \cite{Guichon:1995pu,Guichon:1998xv} for details.  

The particular GP combinations entering the new sum rules are given by:
\bea
\eqlab{GPslope}
&&  P^{\prime\,  (L1, L1)1}(0) \pm  P^{\prime\,  (M1, M1)1}(0) 
 \equiv  \nn\\
 && \qquad \frac{\dd}{\dd \vec q^{\, 2} } \Big[
 P^{ (L1, L1)1}(\vec q^{\, 2} ) \pm  P^{ (M1, M1)1}(\vec q^{\, 2} )  \Big]_{\vec q^{\, 2} =0}.
 \eea
The other properties of the target involved in the sum rules are:
the mass $M$, anomalous magnetic moment $\varkappa$,
the squared Pauli radius $\langle r_2^2\rangle$, two of the four lowest-order spin polarizabilities 
 $\gamma_{E1 M2}$, $\gamma_{E1 E1}$ measured via the real CS process,  and two of the quantities from VVCS:  the slope of the GDH integral $I_1$
and the polarizability $\de_{LT}$. The latter two quantities can be defined in terms 
of the inelastic spin structure functions $g_{1,2}(x,Q^2)$:
\begin{subequations}
\bea
\label{I1sr}
I_1(Q^2) & = & \frac{2M^2}{Q^2}\!\int_0^{x_0}\! \dd x\, g_1(x,Q^2)\, ,\\
\delta_{LT}\,(Q^2) 
&\;=\;& \frac{16\al M^2}{Q^6}\!\int_{0}^{x_0}\! \dd x \, x^2 
[ g_1+g_2](x,Q^2) .
\label{deltaLT}
\eea
\end{subequations}
with $x$ the Bjorken variable, and  $x_0$ indicating the inelastic threshold; $\alpha  \simeq 1/137$ is the fine-structure constant.
With these definitions, the sum rules read:
\begin{widetext}
\begin{subequations}
\bea
\al I_1^\prime(0) &=&  \boxfrac{1}{12} \al \varkappa^2  \langle r_2^2 \rangle 
+  \boxfrac{1}{2} M^2 \gamma_{E1 M2} 
- \boxfrac{3 }{2}  \al M^3\, \big[ P^{\prime(M1, M1)1} (0)+  P^{\prime (L1, L1)1}(0) \big]
\,, 
\label{S1sr3}
\\
\delta_{LT}(0) &=& 
- \gamma_{E1E1} 
+ 3 \al M \, \big[ P^{\prime(M1, M1)1} (0)-  P^{\prime (L1, L1)1}(0) \big] . 
\label{S2sr3}
\eea
\label{qsqrgdhsr}
\end{subequations}
\end{widetext}
The integrals on the {\em lhs}, which sum over the excitation spectrum,
 justify us calling these relations `sum rules'. 
Note that every quantity in each of these relations is observed in a different process:
e.g., the electromagnetic radius and $I_1$ --- in respectively the 
elastic and inelastic electron scattering,  
while the spin polarizabilities and the GPs --- in respectively the RCS and VCS processes. All of these quantities for the nucleon case, besides $\kappa$ which is well known, are currently studied experimentally at electromagnetic beam facilities such as
MAMI, Jefferson Lab, and HI$\gamma$S, see e.g.\ Ref.~\cite{Downie:2011mm} and references therein. The sum rules will provide 
model-independent relations among the experiments that explore the low-energy spin structure of the nucleon.  
Before examining the implications of these sum rules any further,
we briefly explain how they arise. 

Consider the forward VVCS process on a spin-$\nicefrac{1}{2}$ particle (hereby, the nucleon):
$\gamma^\ast (q) + N(p) \to \gamma^\ast(q) + N(p)$,
which is described by two kinematical variables: the photon virtuality 
$q^2 \equiv -Q^2$, and the energy invariant $ \nu \equiv p\cdot q / M $ or the Bjorken variable $x = Q^2 / (2 M \nu)$. The tensor structure of the corresponding scattering amplitude is well-known to decompose
into four scalar amplitudes,  two of which are independent of the nucleon spin, and two 
spin-dependent~\cite{Drechsel:2002ar}. The spin-dependent 
tensor structure reads:
\bea
\label{vvcs}
&&M^{\mu \nu}_{\rm spin}(\nu,Q^2) =
\frac{i}{M}\,\epsilon^{\mu\nu\alpha\beta}\,q_{\alpha}
s_{\beta}\, S_1(\nu, Q^2) \nonumber \\
&&\hspace{1cm}+ \frac{i}{M^3}\,\epsilon^{\mu\nu\alpha\beta}\,q_{\alpha}
(p\cdot q\ s_{\beta}-s\cdot q\ p_{\beta})\, S_2 (\nu, Q^2), \quad \quad 
\eea
where $\nu, \mu$ are the four-vector indicies of the incoming and outgoing photon, $s^\alpha$ is the nucleon covariant
spin vector satisfying $s \cdot p$ = 0, $s^2 = -1$, and $S_1, S_2$ are the spin-dependent 
VVCS amplitudes; $\epsilon_{0123} = +1$.   
The optical theorem relates the imaginary parts of these  
amplitudes to the conventionally defined spin structure functions $g_1$ and $g_2$ 
of inclusive electron-nucleon scattering as:
%
$\im S_1 = (\pi \alpha/\nu) \, g_1$,
$\im S_2  =  (\pi \alpha M/\nu^2) \, g_2$.
Combining it with analyticity and crossing symmetry one can infer the following
dispersion relations (DRs), 
\bea
\label{eq:S1dr}
\re S_1(\nu, Q^2) &=& S_1^{{\rm pole}}(\nu, Q^2)  \nonumber \\
&+& 2 \alpha\,\fint_{\nu_0}^{\infty}  \dd \nu' 
\frac{1}{\nu^{\prime \, 2} - \nu^2} \, g_1(x^\prime, Q^2), \\
\nu \re S_2(\nu, Q^2) &=& \nu  S_2^{{\rm pole}}(\nu, Q^2)  \nonumber \\
&+& 2 M \alpha\,\fint_{\nu_0}^{\infty}  \dd \nu' 
\frac{1}{\nu^{\prime \, 2} - \nu^2} \, g_2(x^\prime, Q^2), 
\label{eq:S2dr}
\eea
 with the first term on the {\it rhs} given by the nucleon pole contribution:
\begin{subequations}
\bea
  S_1^{{\rm pole}} (\nu, Q^2)& = &
 -  \frac{\alpha}{2M} \frac{Q^2}{\nu^2-\nu_B^2}\,F_D(Q^2) G_M(Q^2)  , \\
 \nu S_2^{{\rm pole}} (\nu, Q^2)& = &
 \frac{\alpha}{2} \frac{\nu_B^2}{\nu^2-\nu_B^2}\,F_P(Q^2) G_M(Q^2) ,
\eea\label{pole}
\end{subequations}
where $\nu_B \equiv Q^2 / (2 M)$, $G_M\,(Q^2) = F_D\,(Q^2)+F_P(Q^2)$,  
and $F_{D,P}$ are the nucleon's Dirac and Pauli form factors (FFs).
The second term on the {\it rhs} of Eq.~(\ref{eq:S1dr}) corresponds with the dispersive 
integral over the inelastic states starting from the inelastic threshold $\nu_0$, 
with $x^\prime \equiv Q^2 / (2 M \nu^\prime)$ as the first argument of $g_1$. The slashed integral here denotes the integration in the principal-value sense. 

To derive the sum rules of Eq.~(\ref{qsqrgdhsr}), we look at a low-energy expansion (LEX) 
of  $S_i$ around $\nu = 0$ and $Q^2=0$, for fixed $x\in [0,1]$. 
By using the LEXs of the real and virtual Compton amplitudes as 
detailed in~\cite{DKK98}, for the non-pole part of $S_1$ we obtain:
\bea
\label{S1lex}
(S_1 - S_1^{\rm pole})(\nu, Q^2) =
- \frac{\alpha}{2 M } \,  \varkappa^2 
+   M  \gamma_0 \,  \nu^2  \,+\,  M \ga_L \, Q^2 .
\eea
The well-known $\nu^2$ dependent term in Eq.~(\ref{S1lex}) is proportional to the forward spin polarizability $\ga_0$. The $Q^2$ dependent term is found hereby in terms of the quantities described around \Eqref{GPslope}:
\bea
\gamma_L &=& \frac{\alpha}{6 M^2 } \, \varkappa^2 \, \langle r_2^2 \rangle \,  + \gamma_{E1 M2}  
\nonumber \\
&-& 3 M \alpha \big[ P^{\prime(M1, M1)1} (0)+  P^{\prime (L1, L1)1}(0) \big]. 
\label{cQ}
\eea  
The first term originates from the 
the difference between the Born and pole amplitudes,
which needs to be taken into account since 
the polarizabilities are conventionally defined to affect the non-Born Compton amplitudes. Using,
\beq
F_P(0) =\varkappa,\quad F_P'(0) = - \mbox{$\frac{1}{6}$} \, \varkappa \, \langle r_2^2 \rangle,
\eeq
this difference results in the first terms on the {\it rhs} of Eqs.~(\ref{S1lex}),
 (\ref{cQ}). 

In the final step of the derivation, we expand the {\it rhs} of Eq.~(\ref{eq:S1dr})  
in the photon energy $\nu$, and match the terms
 with the same powers of $\nu^2$ and $Q^2$ in Eq.~(\ref{S1lex}). 
 The $\nu^2$ independent term at $Q^2 = 0$, results in 
the celebrated GDH sum rule~\cite{Gerasimov:1965et,Drell:1966jv}, relating the first moment of $g_1$, defined in Eq.~(\ref{I1sr}),  
to the nucleon's anomalous magnetic moment as: $I_1(0) = - \varkappa^2 / 4$.  
In an analogous way, one obtains from matching the term proportional to $\nu^2$ at $Q^2 = 0$ the  
forward spin polarizability $\ga_0$ sum rule~\cite{GellMann:1954db}. 
The sum rule of Eq.~(\ref{S1sr3}) is obtained by taking the derivative  of Eq.~(\ref{eq:S1dr}),
over $Q^2$ at $Q^2 = 0$, $\nu=0$, and using Eqs.~(\ref{I1sr}), (\ref{S1lex}), and (\ref{cQ}).

In the case of $S_2$ it suffices to let $Q^2=0$.
The LEX of the non-pole part, 
 up to terms of $\cO(\nu^4)$,  yields:
\bea
\label{S2lex}
&& \nu (S_2 - S_2^{\rm pole})(\nu, 0) = \half \al \varkappa G_M(0) -M^2 \nu^2  \big[ \ga_0 +\ga_{E1E1}   \nn\\
&& \quad - 3 M \alpha \big( P^{\prime(M1, M1)1} (0)-  P^{\prime (L1, L1)1}(0) \big)\big],
\eea
where the first term on the {\em rhs} originates again from the difference between the Born and pole contributions. 
Expanding now the DR in Eq.~(\ref{eq:S2dr}), at $\cO(\nu^0)$ we obtain the 
Burkhardt-Cottingham sum rule: 
$\int_{0}^{1}\dd x\,  g_{2}\,(x,\,Q^2)  = 0 $, whereas at $\cO(\nu^2)$
we obtain the  sum rule in Eq.~(\ref{S2sr3}). In the latter step, we
employed the following relation~\cite{Drechsel:2002ar}:
\beq
\lim_{Q^2\to 0}\frac{16M^2\al}{Q^6} \int_{0}^{1}\dd x\, x^2\,  g_{2}\,(x,\,Q^2)
= \de_{LT}-\ga_0.
\eeq

Coming to the empirical implications of the newly proposed relations, Eq.~(\ref{qsqrgdhsr}),
we focus on the proton, and express the poorly
known quantities (here, the polarizabilities)  in terms of the better known quantities. The proton anomalous magnetic
moment is accurately known (we use $\varkappa_p \simeq  1.793$), while
the Pauli radius can be expressed in terms of well-known electric and magnetic radii:
\bea
 \langle r_{2}^2 \rangle &= &
 3/(2M^2) + \big( 1+ \varkappa^{-1} \big)  \langle r_{M}^2 \rangle
 - \varkappa^{-1} \langle r_{E}^2 \rangle \nn\\
 &=& 0.58(4) \;  {\mathrm{fm}}^2
\eea
where we have substituted the recent experimental values for the proton electric and magnetic radii~\cite{Bernauer:2013tpr}.

The polarizabilities receive in principle a $\pi^0$-pole contribution. The following
quantities entering our sum rules are affected this way, see e.g.~\cite{Drechsel:2002ar}:
\begin{eqnarray}
&& \gamma_{E1 E1} = \gamma_{E1 M2} = \alpha C_\pi/2, \quad P^{\prime (M1, M1)1}(0) = C_\pi/6 M,    \nn
\end{eqnarray}
with $C_\pi = g_A / (4 \pi^2 f_\pi^2 m_\pi^2)$. By inserting these values
into the sum rules, Eq.~(\ref{qsqrgdhsr}), one sees the $\pi^0$-pole contribution cancels out.  
We therefore focus on the non-pion-pole part of the polarizabilities alone.
 
The non-pion-pole parts of the spin and generalized polarizabilities have been estimated phenomenologically using a model based on unsubtracted DRs~\cite{Drechsel:2002ar}. 
To indicate the uncertainty due to the phenomenological input of the DR approach, we show in Table~\ref{sumruletest} the DR results obtained using  
MAID2000~\cite{Drechsel:1998hk}  and MAID2007~\cite{Drechsel:2007if}  as input. Combining
these values with the precise values for the proton anomalous magnetic moment and
Pauli radius, the sum rules in Eq.~(\ref{qsqrgdhsr}) yield the predictions for $I_1'(0)$ and $\de_{LT}$ shown
in the two bottom rows of the table. These predictions compare rather favorably with
the available empirical information on these quantities seen in the last column.

\begin{table}[bt]
{\centering 
\begin{tabular}{|c|c|c|c|}
\hline
&  DR & DR &  Experiment \\
&  MAID2000 & MAID2007&  \\
\hline
$ \langle r_2^2 \rangle$  \, [fm$^2$]  & & & $ 0.58 \pm 0.04 $   \\
&  & &  \cite{Bernauer:2013tpr}  \\[1mm]
$\gamma_{E1 M2}$ \, [$10^{-4}$ fm$^4$]  &  $-0.03$  & $-0.11$  & $-0.7 \pm 1.2$   \\
& \cite{Holstein:1999uu} & \cite{Drechsel:2002ar,Drechsel:2007if} & \cite{Martel:2014pba}  \\[1mm]
$\gamma_{E1 E1}$ \, [$10^{-4}$ fm$^4$]  
&  $-4.3$   &  $-4.3$   
& $-3.5 \pm 1.2$   \\
& \cite{Holstein:1999uu} & \cite{Drechsel:2002ar,Drechsel:2007if} & \cite{Martel:2014pba}  \\[1mm]
$P^{\prime (M1, M1)1}(0)$  \, [GeV$^{-5}$]  & 
$-5.8$   & $-4.7$   &$ -$ \\
& \cite{Pasquini:2001yy} & \cite{Pasquini:2001yy} &  \\[1mm]
$P^{\prime (L1, L1)1}(0)$  \, [GeV$^{-5}$]  & 
$3.4$  & $4.3$   & $- $\\
& \cite{Pasquini:2001yy} & \cite{Pasquini:2001yy} & \\
\hline
\quad $I_1^\prime(0)$  \, [GeV$^{-2}$]  &  $6.8$ (SR)  &  $4.0$ (SR)  &   
$7.6 \pm 2.5$   \\
& & & \cite{Prok:2008ev}    \\[1mm]
$\delta_{LT}$   \, [$10^{-4}$ fm$^4$]    \quad 
&   $1.45$   (SR)
&  $1.51$  (SR)
&   $1.34$ (MAID)     \\
& & & \cite{Drechsel:2000ct,Drechsel:2007if}   \\
\hline
\end{tabular}\par
}
\caption{Empirical verification of sum rules  for the  proton. 
First two columns contain the polarizabilities obtained in the DR approach  with respectively MAID2000~\cite{Drechsel:1998hk}  and  MAID2007~\cite{Drechsel:2007if} as input,
and the last two rows give the corresponding sum rule values, to be compared
with experiment in the last column.}
\label{sumruletest}
\end{table}

\onecolumngrid

 \begin{figure}[bth]
\centering
\begin{minipage}[b]{0.45\textwidth}
\includegraphics[width=0.95\textwidth]{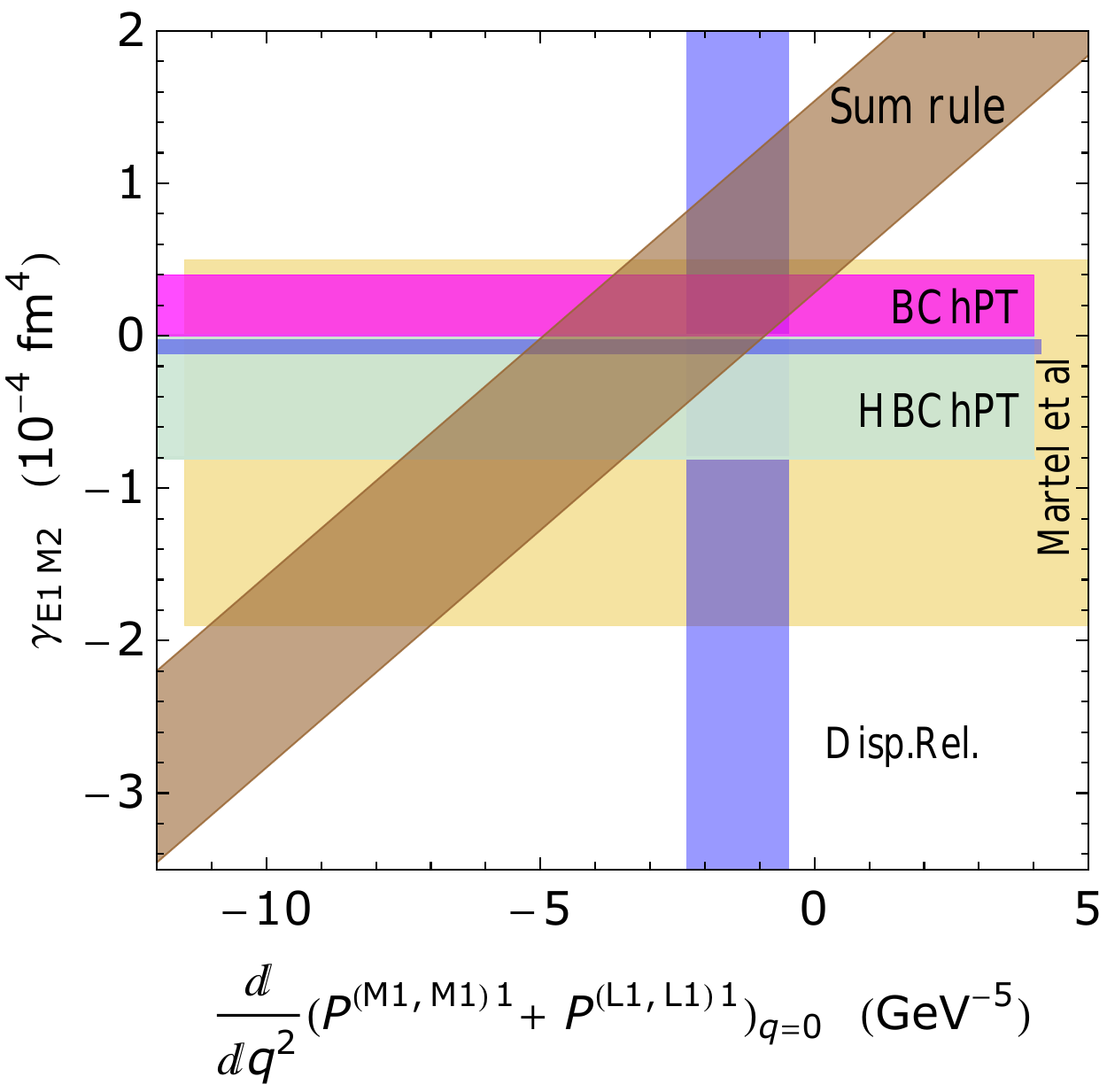}
\end{minipage}
\begin{minipage}[b]{0.45\textwidth}
\includegraphics[width=0.95\textwidth]{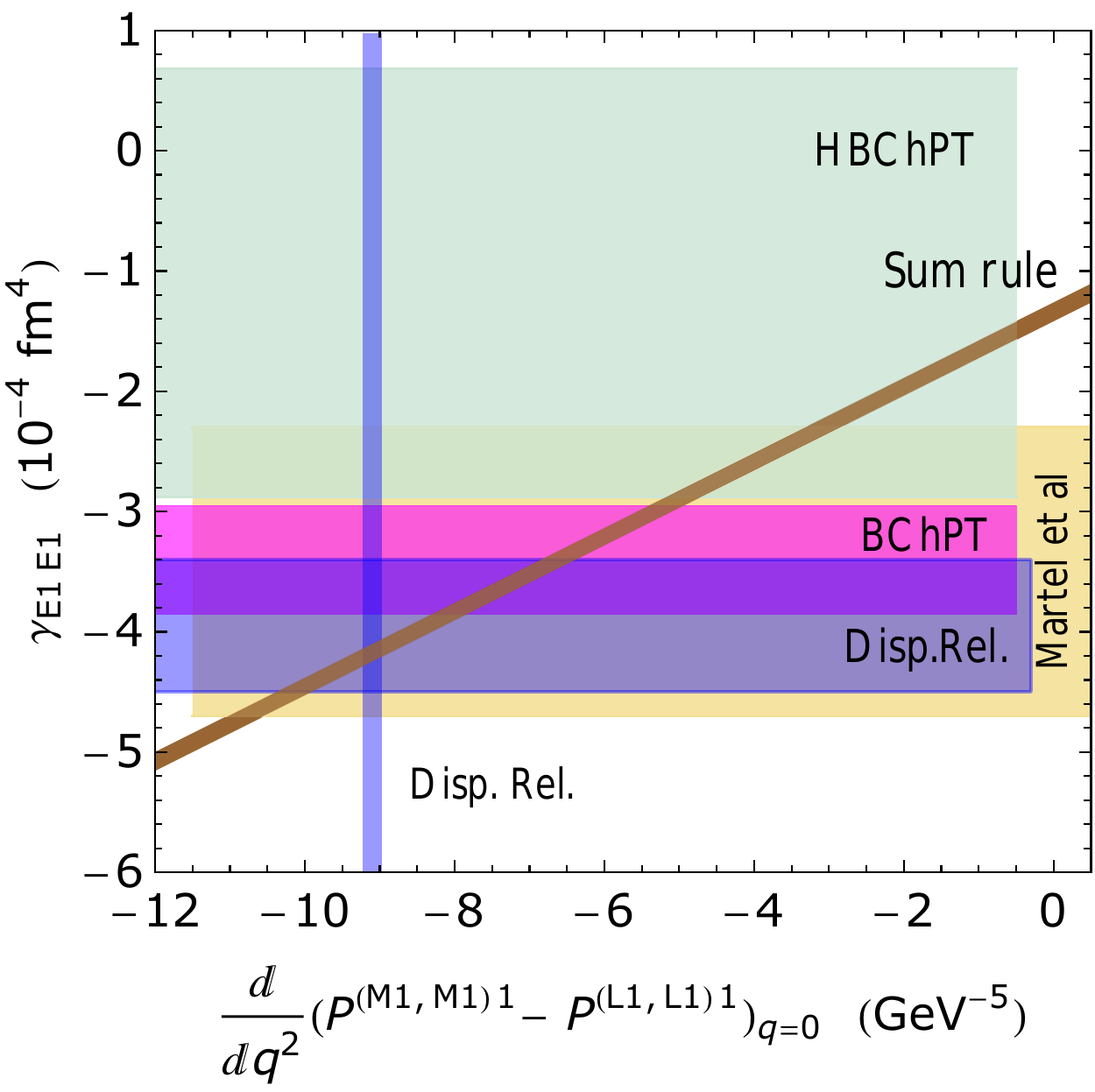}
\end{minipage}
\caption{Spin polarizabilities 
$\gamma_{E1 M2}$ (left panel) and $\gamma_{E1 E1}$ (right panel) 
versus the slope of the VCS GPs for the proton. The (brown) band across the plot is the present sum rule constraint based on the empirical information for $I_1^\prime(0)$,
$ \langle r_2^2 \rangle$ and $\de_{LT}$, cf.\ Table~\ref{sumruletest}. The horizontal beige band is the result of polarized RCS experiments~\cite{Martel:2014pba}. The horizontal light-green band is the empirical extraction based on HBChPT \cite{McGovern:2012ew}.
The horizontal purple band is ChPT prediction~\cite{Lensky:2014efa}. The vertical and horizontal blue bands are the DR evaluations for both the RCS and VCS polarizabilities, cf.\ Table~\ref{sumruletest}. }
\figlab{across}
\end{figure}
\twocolumngrid

Using only the empirical information for $I'(0)$ and $\de_{LT}$, the sum rules provide a slanted 
band
in the plots of $\ga_{E1M2}$ and  $\ga_{E1E1}$  
versus the slopes of the GPs, displayed in~\Figref{across}.
The pioneering experimental values for the spin polarizabilities obtained  recently by the A2 Collaboration at MAMI~\cite{Martel:2014pba}, are given in the table and shown by the broad horizontal (beige-colored) band in the figures labeled `Martel et al'. 
The region where the two bands overlap yields a prediction for the slopes of the GPs. 
A measurement of GP slopes using VCS is required to verify this prediction.

The figures also show the horizontal bands obtained in 
two variants of chiral perturbation theory (ChPT) --- heavy-baryon (HBChPT)
\cite{McGovern:2012ew} and covariant (BChPT)
\cite{Lensky:2014efa,Lensky:2009uv}. The DR estimate of Pasquini
{\em et al.}~\cite{Drechsel:2002ar} for the RCS and VCS polarizabilities, given in the table, is shown in~\Figref{across} by respectively 
the horizontal and vertical blue bands. Again, within the uncertainties, 
the dispersive results are seen to be consistent with the sum rule constraints:
 the vertical, horizontal blue bands and the slanted brown band have a common overlap. Further ChPT calculations of GPs are needed to perform a similar consistency cross check in ChPT.

In conclusion, we have presented two sum rules, Eq.~(\ref{qsqrgdhsr}),
which extend the celebrated Gerasimov-Drell-Hearn and Burkhardt-Cottingham sum rules, respectively. The new sum rules,
 involve low-energy electromagnetic properties which are accessed in different experiments: the Pauli radius of the target (e.g., nucleon),  spin polarizabilities, and the slopes of two of its four lowest order generalized polarizabilities. The present empirical and phenomenological information on these quantities for the proton is shown
to be consistent with the sum rules, albeit with large experimental uncertainties. 
New experiments, ongoing at the 
MAMI, Jefferson Lab, and HI$\gamma$S facilities, will deliver a substantially improved  input of the quantities entering this sum rule, and thus provide a new model independent  
constraint on the low-energy spin structure of the proton.  It will also be interesting to examine the sum rules in theory. They can cross check the
consistency of different variants of chiral perturbation theory, as all of the involved quantities can in principle be calculated to a given order in the chiral expansion. In a broader context, 
the sum rules establish a fundamental relation between the low-momentum transfer light absorption and scattering 
on a polarized spin-1/2 target.

\section*{Acknowledgements}
We would like to thank Barbara Pasquini, Chungwen Kao, Philippe Martel, and Karl Slifer for helpful discussions.  This work was supported by the Deutsche Forschungsgemeinschaft (DFG) 
through the Collaborative Research Center [The Low-Energy Frontier of the Standard Model (SFB 1044)] and the Cluster of Excellence [Precision Physics, Fundamental
Interactions and Structure of Matter (PRISMA)].

\end{document}